\documentclass{article}
\usepackage[a4paper]{geometry}
\usepackage{amsmath,amssymb}
\usepackage{epsfig}
\usepackage{psfrag}

\usepackage{bm}

\def\erre{{\bf r}}
\def\erredot{\dot{\bf r}}
\def\bu{{\bf u}}
\def\bv{{\bf v}}
\def\bq{{\bf q}}
\def\bqdot{\dot{\bf q}}
\def\alphadot{\dot{\alpha}}
\def\deltadot{\dot{\delta}}
\def\erho{{\bf e}^\rho}
\def\ealpha{{\bf e}^\alpha}
\def\edelta{{\bf e}^\delta}

\def\rhodot{\dot{\rho}}
\def\etabf{\bm{\eta}}

\def\DD{{\bf D}}
\def\EE{{\bf E}}
\def\FF{{\bf F}}
\def\GG{{\bf G}}
\def\JJ{{\bf J}}

\def\angmom{{\bf c}}
\def\energy{{\cal E}}
\def\lenz{{\bf L}}

\def\bDelta{\bm{\Delta}}
\def\bxi{\bm{\xi}}
\def\bXi{\bm{\Xi}}

\def\bzero{{\bf 0}}

\def\Avec{{\bf A}}
\def\Att{{\cal A}}
\def\Rvec{{\bf R}}
\def\bPhi{\bm{\Phi}}
\def\Rcal{\mathcal{R}}
\def\bPsi{\bm{\Psi}}
\def\R{\mathbb{R}}
\def\C{\mathbb{C}}
\def\N{\mathbb{N}}

\def\oldu{\mathfrak{u}}
\def\newu{\mathfrak{v}}

\newtheorem{remark}{\bf Remark}
\newtheorem{lemma}{\bf Lemma}
\newtheorem{proposition}{\bf Proposition}
\newtheorem{theorem}{\bf Theorem}
\newtheorem{corollary}{\bf Corollary}

\begin{document}

\title{\bf Keplerian integrals, elimination theory and identification
  of very short arcs in a large database of optical observations}

\author{Giovanni~F. Gronchi
\footnote{Dip. di Matematica, Univ. di Pisa, 
Italy
     {\tt gronchi@dm.unipi.it}}\;, 
Giulio Ba\`u
\footnote{Dip. di Matematica, Univ. di Pisa, 
Italy
     {\tt giulio.bau@unipi.it}}\;,
Andrea Milani
 \footnote{Dip. di Matematica, Univ. di Pisa, 
Italy
     {\tt milani@dm.unipi.it}}
}

 
\maketitle

\begin{abstract}
  The modern optical telescopes produce a huge number of asteroid
  observations, that are grouped into very short arcs (VSAs), each
  containing a few observations of the same object in one single night.  To
  decide whether two VSAs, collected in different nights, refer to the same
  observed object we can attempt to compute an orbit with the observations of
  both arcs: this is called the {\em linkage} problem. Since the number of
  orbit computations to be performed is very large, we need efficient methods
  of orbit determination.  Using the first integrals of Kepler's motion we can
  write algebraic equations for the linkage problem, which can be put in
  polynomial form, see \cite{gdm10}, \cite{gfd11}, \cite{gbm15}.  The
  equations introduced in \cite{gbm15} can be reduced to a univariate
  polynomial of degree 9: the unknown is the topocentric distance $\rho$ of
  the observed body at the mean epoch of one of the VSAs.  Using elimination
  theory we show an optimal property of this polynomial: it has the least
  degree among the univariate polynomials in the same variable that are
  consequence of the algebraic conservation laws and are obtained without
  squaring operations, that can be used to bring these algebraic equations in
  polynomial form.  
  In this paper we also introduce a procedure to join three VSAs belonging to
  different nights: from the conservation of angular momentum at the three
  mean epochs of the VSAs, we obtain a univariate polynomial equation of
  degree 8 in the topocentric distance $\rho_2$ at the intermediate epoch.
  This algorithm has the same computational complexity as the classical method
  by Gauss, but uses more information, therefore we expect that it can produce
  more accurate results. These results can be used as better preliminary
  orbits to compute a least squares orbital solution with three VSAs.
  For both methods, linking two and three VSAs, we also discuss how to select
  the solutions, making use of the full two-body dynamics, and show some
  numerical tests comparing the results with the ones obtained by Gauss'
  method.


\end{abstract}

\section{Introduction}

We consider
very short arcs (VSAs) of optical observations of a solar system body whose
motion is dominated by the gravitational attraction of the Sun. These small
sets of observations are called {\em tracklets} and the corresponding
arc described in the sky is usually too short to compute a least squares
orbit.
In each observing night we can detect thousands of these data, so that
it is difficult to decide whether two such arcs, collected in different
nights, refer to the same observed body.  This gives rise to an
identification problem, that can be solved by computing an orbit with
the information contained in two or more tracklets.

Using the classical methods of initial orbit determination, those
by Laplace \cite{laplace} or Gauss \cite{gauss}, we usually cannot
compute a preliminary orbit with three observations belonging to the same
VSA because they are too close in time and the arc is usually too
short.  Even using observations taken from two different VSAs 
it may be difficult to compute an orbit.
Laplace's or Gauss' methods in most cases work well if we use three different
observations from three VSAs. In this case to compute a preliminary orbit we
have to find the roots of a univariate polynomial of degree 8 (see
\cite{mg10}), that correspond to the possible values of the radial distance
(geocentric for Laplace, topocentric for Gauss) of the observed body at a
given epoch (the mean epoch of the observations $\sum_{h=1}^3 t_h/3$ for
Laplace, the central epoch $t_2$ for Gauss).

Assume for simplicity that we deal with this identification problem using the
observations made by a single telescope performing an asteroid survey, like
Pan-STARRS \cite{panstarrs}, or the next generation telescope LSST
\cite{lsst}.
The average number of observations per night is $N \approx 10^4$ for
Pan-STARRS, and presumably we shall have $N \approx 10^5$ for LSST.
%
To perform systematically the identification by Gauss' method using
the data of three observing nights we should test compatibility for
$O(N^3)$ triples of observations. This is clearly a cumbersome task.
The identification of two VSAs is usually called the {\em linkage} problem,
and it has been recently studied in \cite{gdm10}, \cite{gfd11}, \cite{gbm15}
using the first integrals of the two-body motion.

In \cite{gbm15} the authors introduced a univariate polynomial equation of
degree 9 for the linkage problem, which is comparable with the equation of
Gauss' method. This equation is derived in a concise way in
Section~\ref{s:linktwo}.  Moreover, we discuss an optimal property of such
polynomial. Using algebraic elimination theory, we show that it has the least
degree among the univariate polynomials that are consequence of the algebraic
conservation laws of Kepler's problem, provided we drop the dependence between
the inverse of the heliocentric distance $1/|\erre|$ appearing in the
Keplerian potential and the topocentric distance $\rho$. This approach avoids
the squaring operations needed in \cite{gdm10}, \cite{gfd11} to bring the
selected equations\footnote{In these papers not all the algebraic conservation
  laws are used.} into a polynomial form.
In Section~\ref{s:comptwo} we sketch a method to check the validity of the
identification and select solutions according to some compatibility conditions,
similar to the ones in \cite{gdm10}, that use the full two-body dynamics.

An orbit computed with two VSAs is usually not as reliable as one computed
with three observations, each picked up in a different VSAs, because the
latter usually represents a longer arc. To obtain more reliable results we
have to join together at least three VSAs.
In Section~\ref{s:linkthree} we introduce a univariate polynomial equation of
degree 8 to link three VSAs of optical observations by means of the
conservation of angular momentum only.  Then the other laws of Kepler's motion
can be used to set up restrictive compatibility conditions, allowing us to
test the identification and select solutions.

Assume we set up an identification procedure with a large database of asteroid
observations. For simplicity, we can consider three observing nights, in which
we collect $O(N)$ VSAs of observations per night.  We can try to identify
pairs of VSAs belonging to the first two nights by applying $O(N^2)$ times the
linkage algorithm introduced in \cite{gbm15} and reviewed in
Section~\ref{s:linktwo}.  The output is composed by preliminary orbits
obtained with pairs of VSAs. If the thresholds in the controls for acceptance
(see Section~\ref{s:comptwo}) are well selected, we do not obtain more than
$O(N)$ pairs of VSAs, in fact the number of different objects observed in the
two nights is $O(N)$.  Then we can apply the method to link three VSAs
introduced in Section~\ref{s:linkthree} to the $O(N)$ selected pairs and the
$O(N)$ VSAs of the third observing night.
We conclude that this identification problem can be faced by $O(N^2)$
computations of roots of a polynomial of degree 9 or 8, instead of $O(N^3)$
computations of roots of Gauss' polynomial.

\section{Keplerian integrals}
\label{s:kepint}

We consider the Keplerian motion of a celestial body around a center of force,
set at the origin of a given reference system, that in the asteroid case
corresponds to the center of the Sun.  Optical observations of the body are
made by a telescope whose heliocentric position is a known function of time.
Then the heliocentric position and velocity of the body are given by
\begin{equation}
\erre = \rho\erho + \bq, \hskip 1.5cm
\erredot = \rhodot\erho + \rho\etabf + \bqdot,
\label{rrdot}
\end{equation}
where $\bq, \bqdot$ are the heliocentric position and velocity of the
observer, $\rho, \rhodot$ are the topocentric radial distance and velocity,
$\erho$ is the {\em line of sight} unit vector, which can be written in terms
of the topocentric right ascension $\alpha$ and declination $\delta$ as
\[
\erho=(\cos\delta\cos\alpha,\cos\delta\sin\alpha,\sin\delta).
\]
Moreover in (\ref{rrdot}) we set
\[
\etabf = \alphadot\cos\delta\ealpha + \deltadot\edelta,
\]
where
\[
\ealpha = (\cos\delta)^{-1}\frac{\partial \erho}{\partial\alpha},\hskip 1cm
\edelta = \frac{\partial \erho}{\partial\delta},
\]
and $\alphadot, \deltadot$ are the angular rates.
The Keplerian integrals, represented by the angular momentum vector
$\angmom$, the Laplace-Lenz vector
$\mathbf{L}$ and the energy $\mathcal{E}$, are defined by 
\begin{equation}
\angmom = \erre\times\erredot,\qquad
\mu\lenz = \Bigl(|\erredot|^2 - \frac{\mu}{|\erre|}\Bigr)\erre -
(\erre\cdot\erredot)\erredot,\qquad
\energy = \frac{1}{2}|\erredot|^2 - \frac{\mu}{|\erre|},
\label{kepint}
\end{equation}
as functions of $\mathbf{r}, \dot{\mathbf{r}}$. Given the values of
$\alpha,\delta, \alphadot, \deltadot$, they can be written as algebraic
functions of $\rho, \rhodot$ using relations (\ref{rrdot}).

\section{Linking two VSAs}
\label{s:linktwo}

Given a very short arc of optical observations ($\alpha_i$, $\delta_i$),
$i=1\ldots m$, made by the same station at $t_{i}$ different times, it
is often possible to compute the {\em attributable} vector (see \cite{ident4})
\[
\Att=(\alpha,\delta,\alphadot,\deltadot)
\]
at the mean epoch $\bar{t}=\frac{1}{m}\sum_{i=1}^{m}t_{i}$.  The missing
quantities to obtain a preliminary orbit are the topocentric distance and
velocity $\rho$, $\rhodot$ at $t=\bar{t}$. 
When the second derivatives ($\ddot{\alpha}$, $\ddot{\delta}$) are either not
available (if $m=2$), or not accurate enough due to the errors in the
observations, then the attributable summarizes essentially all the information
contained in the VSA.  In this case a preliminary orbit can be obtained by
linking together two different VSAs.

The key idea of the linkage method is to use the conservation of the Keplerian
integrals $\angmom$, $\lenz$, $\energy$ at the two mean epochs $\bar{t}_{1},
\bar{t}_{2}$ of two attributables $\Att_{1}, \Att_{2}$:
\begin{equation}
\angmom_1 = \angmom_2, \qquad
\lenz_1 = \lenz_2,\qquad
\energy_1 = \energy_2,
\label{conslaws}
\end{equation}
where the indexes $1, 2$ refer to the epoch.

Below we derive in a concise way the polynomial equations for the linkage
problem introduced in \cite{gbm15}, and we review the procedure to obtain the
univariate polynomial of degree 9 giving the possible values for the
topocentric distance $\rho_2$.  Moreover, we show here an optimal property of
this polynomial.

\subsection{Conservation of angular momentum}
\label{s:angmom}

The angular momentum as function of $\rho,\rhodot$ can be written as
\[
\angmom(\rho,\rhodot) = \erre \times \erredot = \DD \rhodot + \EE
\rho^2 + \FF \rho + \GG,
\]
where
\[
\begin{array}{l}
\DD = \bq\times\erho,\cr
\EE = \alphadot\cos\delta\erho\times\ealpha +
\deltadot\erho\times\edelta 
= \alphadot\cos\delta\edelta - \deltadot\ealpha ,\cr
\FF = \alphadot\cos\delta\bq\times\ealpha + \deltadot\bq\times\edelta + \erho\times\bqdot,\cr
\GG = \bq\times\bqdot .\cr
\end{array}
\]
Then the equation
\[
\angmom_1=\angmom_2,
\]
representing the conservation of the angular momentum are written as
\begin{equation}
\DD_1\rhodot_1 - \DD_2\rhodot_2 = \JJ(\rho_1,\rho_2),
\label{eq_AM}
\end{equation}
where 
\begin{equation}
\JJ(\rho_1,\rho_2) = \EE_2\rho_2^2 - \EE_1\rho_1^2 + \FF_2\rho_2 -
\FF_1\rho_1 + \GG_2 - \GG_1 .
\label{JJ}
\end{equation}
We can eliminate the radial velocities $\rhodot_1, \rhodot_2$ by
making the scalar product with $\DD_1\times\DD_2$, that gives the quadratic
polynomial
\begin{equation}
q(\rho_1,\rho_2) := \DD_1\times\DD_2\cdot\JJ(\rho_1,\rho_2) = 0
\label{qpoly}
\end{equation}
in the variables $\rho_1, \rho_2$.
The radial velocities are given by
\begin{equation}
\small
\rhodot_1(\rho_1,\rho_2) =
\frac{(\JJ\times\DD_2)\cdot(\DD_1\times\DD_2)}
{|\DD_1\times\DD_2|^2},
\hskip 0.4cm
\rhodot_2(\rho_1,\rho_2) =
\frac{(\JJ\times\DD_1)\cdot(\DD_1\times\DD_2)}
{|\DD_1\times\DD_2|^2}.
\label{rhojdot}
\end{equation}
These expressions are obtained by projecting (\ref{eq_AM}) onto the
vectors $\DD_1\times(\DD_1\times\DD_2)$ and
$\DD_2\times(\DD_1\times\DD_2)$, generating the plane orthogonal to
$\DD_1\times\DD_2$. Therefore
using such expressions of $\rhodot_1, \rhodot_2$ we have
\[
(\angmom_1-\angmom_2)\times(\DD_1 \times \DD_2) = \bzero
\]
whatever the values of $\rho_1,\rho_2$.
%

\subsection{The univariate polynomial $\mathfrak{u}$}

By relations (\ref{rhojdot}) we can eliminate the
dependence on $\rhodot_1, \rhodot_2$ in the Laplace-Lenz
and energy conservation laws 
\begin{equation}
\lenz_1 = \lenz_2, \hskip 1cm
\energy_1 = \energy_2.
\label{energylenz}
\end{equation}
These are algebraic equations in $\rho_1, \rho_2$ that are not polynomial 
because of the terms $1/|\erre_1|, 1/|\erre_2|$.
However, in the equation
\begin{equation}
\bxi := \bigl[\mu(\lenz_1-\lenz_2) - 
(\energy_1\erre_1-\energy_2\erre_2)\bigr]\times(\erre_1 - \erre_2) = \bzero,
\label{bxi}
\end{equation}
which is a consequence of (\ref{energylenz}), the terms
$1/|\erre_1|, 1/|\erre_2|$ cancel out.  
The monomials of $\bxi$ with the highest total degree, i.e. 6, are all
parallel to $\erho_1\times\erho_2$, so that we consider the bivariate
polynomials
\begin{equation}
p_1 = \bxi\cdot\erho_1,\qquad p_2=\bxi\cdot\erho_2
\label{p1p2}
\end{equation}
having total degree 5.
%
In \cite{gbm15} the authors show that the overdetermined bivariate
polynomial system
\[
q = 0, \qquad \bxi = \bzero
\]
is consistent, i.e. its set of solutions in $\C^2$ is not empty, and is
equivalent to
\[
q = p_1 = p_2 = 0.
\]
Moreover, if we consider the resultants (see \cite{cox})
\[
\mathfrak{u}_1 = \mathrm{Res}(p_1,q,\rho_1),\hskip 1cm
\mathfrak{u}_2 = \mathrm{Res}(p_2,q,\rho_1),
\]
then the greatest common divisor of $\mathfrak{u}_1$ and $\mathfrak{u}_2$,
\begin{equation}
  \oldu = 
  \mathrm{gcd}(\mathfrak{u}_1,\mathfrak{u}_2),
\label{oldupol}
\end{equation}
is a univariate polynomial in the variable $\rho_2$ of degree 9.

\begin{remark}
Since in this problem the role of $\rho_1$ and $\rho_2$ is symmetric, for a
generic choice of the data $\Att_j, \bq_j, \bqdot_j, j=1,2$, we obtain an
analogous result by eliminating the variable $\rho_2$, instead of $\rho_1$,
from $p_1, p_2$.
\label{r:symmetric}
\end{remark}

\noindent We also recall the construction used in \cite{gbm15} to compute
$\mathfrak{u}_j$, $j=1,2$. 
We can write
\[
q(\rho_1,\rho_2) = 
\sum_{h=0}^2b_h(\rho_2)\rho_1^h,
\]
where
\[
b_0(\rho_2) = q_{0,2}\rho_2^2 +  q_{0,1}\rho_2 +  q_{0,0},
\qquad
b_1 = q_{1,0},
\qquad
b_2 = q_{2,0},
\]
with the coefficients $q_{h,k}$ depending only on the data
$\Att_j,\bq_j,\bqdot_j$, $j=1,2$.
Moreover, we have
\begin{equation}
p_1(\rho_1,\rho_2) = 
\sum_{h=0}^4a_{1,h}(\rho_2)\rho_1^h,
\hskip 1cm
p_2(\rho_1,\rho_2) =
\sum_{h=0}^5a_{2,h}(\rho_2)\rho_1^h,
\label{eqpj}
\end{equation}
for some univariate polynomials $a_{k,h}$ whose degrees are described
by the upper small circles used to construct Newton's polygons of $p_1,p_2$
in Figure~{\ref{polynewt}.

\begin{figure}[t!]
\psfragscanon
\centerline{\epsfig{figure=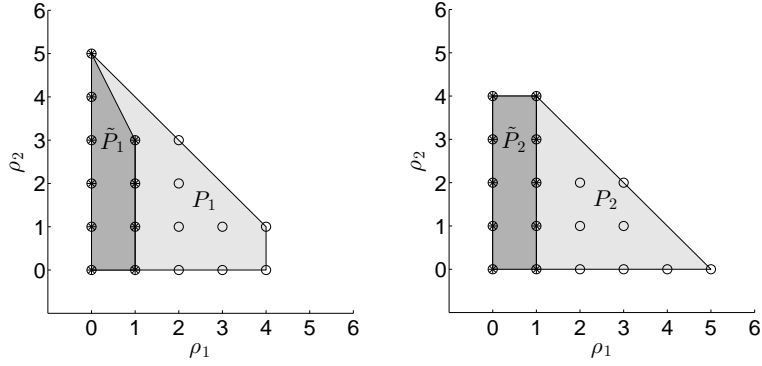,width=12cm}}
\psfrag{Q}{$Q$}\psfrag{P1}{$P_1$}\psfrag{P2}{$P_2$}
\psfrag{TP1}{$\tilde{P}_1$}\psfrag{TP2}{$\tilde{P}_2$}
\psfragscanoff
\caption{We draw Newton's polygons $P_j$, $\tilde{P}_j$ for the polynomials
  $p_j, \tilde{p}_j$, $j=1,2$. In this figure the polygons are overlapping:
  the nodes with circles correspond to the (multi-index) exponents of the
  monomials in $p_j$; the nodes with asterisks correspond to the exponents of
  the monomials in $\tilde{p}_j$.}
\label{polynewt}
\end{figure}
Assume $q_{2,0}, q_{0,2}\neq 0$. From $q=0$
we obtain
\begin{equation}
  \rho_1^h = 
  \beta_h\rho_1 + \gamma_h, \qquad h=2,3,4,5,
\label{rho1h}
\end{equation}
where
\[
\beta_2 = -\frac{b_1}{b_2}, \qquad \gamma_2 = -\frac{b_0}{b_2},
\]
and
\[
\beta_{h+1} = \beta_h\beta_2 + \gamma_h, \qquad \gamma_{h+1} = \beta_h\gamma_2, \qquad h=2,3,4. 
\]
Inserting (\ref{rho1h}) into (\ref{eqpj}) we obtain
\begin{equation}
\tilde{p}_j(\rho_1,\rho_2) = \tilde{a}_{j,1}(\rho_2)\rho_1 + 
\tilde{a}_{j,0}(\rho_2), \hskip 1cm  j=1,2,
\label{pjtilde}
\end{equation}
where
\begin{eqnarray}
&&\tilde{a}_{1,1} = a_{1,1} + \sum_{h=2}^4 a_{1,h}\beta_h,
\hskip 0.5cm 
\tilde{a}_{1,0} = a_{1,0} + \sum_{h=2}^4 a_{1,h}\gamma_h, \label{atilde1}\\
&&\tilde{a}_{2,1} = a_{2,1} + \sum_{h=2}^5 a_{2,h}\beta_h,
\hskip 0.5cm 
\tilde{a}_{2,0} =  a_{2,0} + \sum_{h=2}^5 a_{2,h}\gamma_h.\label{atilde2}
\end{eqnarray}
In Figure~\ref{polynewt} we also draw Newton's polygons of $\tilde{p}_1$,
$\tilde{p}_2$. In this case the nodes with asterisks correspond to the
exponents of the monomials in $\tilde{p}_j$ and the upper asterisks describe
the degrees of the polynomials $\tilde{a}_{k,h}$.

Let us introduce the polynomials
\[
\mathfrak{v}_1 = \mathrm{Res}(\tilde{p}_1,q,\rho_1),
\hskip 1cm
\mathfrak{v}_2 = \mathrm{Res}(\tilde{p}_2,q,\rho_1).
\]
We can show the following result.
\begin{lemma}
By the properties of resultants we find that
\begin{equation}
\mathfrak{u}_1 = q_{2,0}^3 \mathfrak{v}_1, \qquad \mathfrak{u}_2 = q_{2,0}^4
\mathfrak{v}_2.
\label{ujvsvj}
\end{equation}
\end{lemma}
{\em Proof.}
We prove the first relation; the proof of the second one is similar. We have
\[
\mathfrak{u}_1 = \mathrm{Res}(p_1,q,\rho_1)=
\det\left[\begin{array}{cccccc}
a_{10} & 0 & b_{0} & 0 & 0 & 0\\
a_{11} & a_{10} & b_{1} & b_{0} & 0 & 0\\
a_{12} & a_{11} & b_{2} & b_{1} & b_{0} & 0\\
a_{13} & a_{12} & 0 & b_{2} & b_{1} & b_{0}\\
a_{14} & a_{13} & 0 & 0 & b_{2} & b_{1}\\
0 & a_{14} & 0 & 0 & 0 & b_{2}
\end{array}\right].
\]
By performing raw operations and by the properties of determinants we obtain
\begin{eqnarray*}
\mathrm{Res}(p_1,q,\rho_1)&=&
\det\left[\begin{array}{cccccc}
a_{10} &0 &b_{0} &0 &0 &0\\
a_{11}+\gamma_{2}a_{13}+\beta_{2}\gamma_{2}a_{14} &\tilde{a}_{10} &b_{1}&0 &0 &0\\
a_{12}+\beta_{2}a_{13}+(\beta_{2}^{2}+\gamma_{2})a_{14}\hskip 0.1cm &\tilde{a}_{11} &b_{2}&0 &0 &0\\
a_{13}+\beta_{2}a_{14} & a_{12}+\beta_{2}a_{13}+\beta_{3}a_{14} &0  &b_{2} &0 &0\\
a_{14} & a_{13}+\beta_{2}a_{14} &0  &0  &b_{2} &0\\
0 & a_{14} &0  &0  &0  &b_{2}
\end{array}\right] = \\
&=& \det
\left[\begin{array}{cccccc}
\tilde{a}_{10} & 0 & b_{0} & 0 & 0 & 0\\
\tilde{a}_{11} & \tilde{a}_{10} & b_{1} & 0 & 0 & 0\\
0 & \tilde{a}_{11} & b_{2} & 0 & 0 & 0\\
a_{13}+\beta_{2}a_{14}\hskip 0.1cm & a_{12}+\beta_{2}a_{13}+\beta_{3}a_{14} & 0 & b_{2} & 0 & 0\\
a_{14} & a_{13}+\beta_{2}a_{14} & 0 & 0 & b_{2} & 0\\
0 & a_{14} & 0 & 0 & 0 & b_{2}
\end{array}\right]
=
b_2^3\mathrm{Res}(\tilde{p}_1,q,\rho_1).
\end{eqnarray*}
The last matrix is obtained from the previous one by adding to its first 
column a suitable multiple of the third column.

\rightline{$\square$}

\subsection{An optimal property of the polynomial $\oldu$}
\label{s:minim}

If we consider the auxiliary variable $u$ together with the polynomial
relation
\begin{equation}
u^2|\erre|^2 = \mu^2,    
\label{urhodep}
\end{equation}
then the Keplerian integrals introduced in (\ref{kepint}) can be viewed as
polynomials in the variables $\rho,\rhodot,u$.
In particular, we obtain
\[
\lenz = (|\erredot|^2 - u)\erre - (\erredot\cdot\erre)\erredot, 
\hskip 1cm
\energy = \frac{1}{2}|\erredot|^2 - u.
\]
We observe that, for all $\rho,\rhodot,u$,
\begin{equation}
\angmom\cdot\lenz = 0,
\hskip 1cm
\mu^2|\lenz|^2 = u^2|\erre|^2 + 2\energy|\angmom|^2;
\label{intrelgen}
\end{equation}
the second relation generalizes the classical formula relating eccentricity,
energy and angular momentum.  

The full polynomial system 
\begin{equation}
\angmom_1=\angmom_2,\qquad \mu\lenz_1=\mu\lenz_2,\qquad
\energy_1=\energy_2,\qquad u^2_j|\erre_j|^2 = \mu^2 \quad (j=1,2),
\label{fullsys}
\end{equation} 
with unknowns $\rho_1, \rho_2, \rhodot_1, \rhodot_2, u_1, u_2$,
is generically not consistent, see Corollary~\ref{inconsist} at the end of
this section. 
Next we show that, if we drop the dependence between $u_j$ and $\rho_j$ given
by relation (\ref{urhodep}),
we obtain a consistent polynomial system, and the univariate polynomial
$\oldu$ of degree 9 introduced in \cite{gbm15} has the least degree among the
polynomials in $\rho_2$ that are a consequence of the polynomials in
(\ref{conslaws}).
Therefore $\oldu$ has the least degree among the polynomials in $\rho_2$,
consequences of the algebraic Keplerian integrals and obtained without
squaring operations, which can be used to bring the algebraic conservation laws
in polynomial form.
%

Let
\[
I\subseteq \R[\rho_1,\rho_2,\rhodot_1,\rhodot_2,u_1,u_2]
\] 
be the ideal of the polynomial ring in the variables
$\rho_1,\rho_2,\rhodot_1,\rhodot_2,u_1,u_2$, with real coefficients,
generated by the seven polynomials
\[
\angmom_1-\angmom_2,\qquad \mu\lenz_1-\mu\lenz_2,\qquad \energy_1-\energy_2,
\]
where we write $u_j$ in place of  $\mu/|\erre_j|$ for $j=1,2$.

We recall that a set $\{\mathfrak{g}_1,\ldots,\mathfrak{g}_n\}$, with
$n\in\N$, is a Groebner basis of a polynomial ideal $I$ for a fixed monomial
order $\succ$ if and only if the leading term (for that order) of any element
of $I$ is divisible by the leading term of one $\mathfrak{g}_j$, see
\cite{cox}.
The main result of this section is the following.
\begin{theorem}
  For a generic choice of the data $\Att_j,\bq_j,\bqdot_j$, $j=1,2$, we can
  find a set of polynomials
\[
\mathfrak{g}_1\ldots\mathfrak{g}_6 \in \R[\rho_1,\rho_2,\rhodot_1,\rhodot_2,u_1,u_2],
\]
which is a Groebner basis of the ideal $I$ for the
lexicographic order
\begin{equation}
\rhodot_1\succ \rhodot_2\succ u_1\succ u_2 \succ \rho_1 \succ \rho_2,
\label{lexord}
\end{equation}
such that
\[
\mathfrak{g}_6 = \mathfrak{u},
\]
where $\oldu$ is the polynomial defined in (\ref{oldupol}).
\label{mainteo}
\end{theorem}

{\em Proof.}
Assuming 
\[
\DD_1\times\DD_2\neq \bzero, \hskip 1cm 
\erho_1\times\erho_2\neq \bzero, 
\]
we consider the following 
set of generators of the ideal $I$:
\begin{eqnarray*}
\mathfrak{q}_1 &=& (\angmom_1-\angmom_2)\cdot\DD_1\times\DD_2,\\
\mathfrak{q}_2 &=& (\angmom_1-\angmom_2)\cdot\DD_1\times(\DD_1\times\DD_2),\\
\mathfrak{q}_3 &=& (\angmom_1-\angmom_2)\cdot\DD_2\times(\DD_1\times\DD_2),\\
\mathfrak{q}_4 &= &\mu(\lenz_1-\lenz_2)\cdot\erho_1\times\erho_2,\\
\mathfrak{q}_5 &= &\mu(\lenz_1-\lenz_2)\cdot\DD_1,\\
\mathfrak{q}_6 &= &\mu(\lenz_1-\lenz_2)\cdot\DD_2,\\
\mathfrak{q}_7 &=& \energy_1-\energy_2.
\end{eqnarray*}
%
The first three polynomials have the form
\begin{eqnarray*}
\mathfrak{q}_1 &=& q,\\
\mathfrak{q}_2 &=& |\DD_1\times\DD_2|^2\rhodot_1 - 
\JJ\cdot \DD_1\times(\DD_1\times\DD_2),\\
\mathfrak{q}_3 &=&  |\DD_1\times\DD_2|^2\rhodot_2 - 
\JJ\cdot \DD_2\times(\DD_1\times\DD_2),
\end{eqnarray*}
with $q= q(\rho_1,\rho_2)$, $\JJ = \JJ(\rho_1,\rho_2)$ defined in (\ref{qpoly}), (\ref{JJ}) respectively.
The other generators can be written as  
\begin{eqnarray*}
\mathfrak{q}_4 &=& -(\DD_1\cdot\erho_2)u_1 -(\DD_2\cdot\erho_1)u_2 + \mathfrak{f}_4,\\
\mathfrak{q}_5 &=& (\DD_1\cdot\erre_2)u_2 + \mathfrak{f}_5,\\
\mathfrak{q}_6 &=& -(\DD_2\cdot\erre_1)u_1 + \mathfrak{f}_6,\\
\mathfrak{q}_7 &=& -u_1 + u_2 + \mathfrak{f}_7,
\end{eqnarray*}
for some polynomials $\mathfrak{f}_j =
\mathfrak{f}_j(\rho_1,\rho_2,\rhodot_1,\rhodot_2)$, $j=4\ldots 7$.
Set
\[
A = \DD_1\cdot\erho_2+\DD_2\cdot\erho_1 =
(\bq_1-\bq_2)\cdot\erho_1\times\erho_2.
\]
Assuming the three terms
\[
A, \quad \DD_2\cdot\erho_1,\quad \DD_1\cdot\erho_2
\]
do not vanish, we can substitute the generators
$\mathfrak{q}_4\ldots\mathfrak{q}_7$ with the polynomials
\begin{eqnarray*}
\mathfrak{p}_4 &=& (\DD_1\cdot\erho_2)\mathfrak{q}_{7}-
  \mathfrak{q}_{4} = A u_2 + \mathfrak{a}_1,\\
\mathfrak{p}_5 &=& 
-(\DD_2\cdot\erho_1)\mathfrak{q}_{7}-
\mathfrak{q}_{4}
= A u_1 + \mathfrak{a}_2,\\
\mathfrak{p}_6 &=& 
(\DD_1\cdot\erre_2)\mathfrak{p}_{4}-
A\mathfrak{q}_5,\\
\mathfrak{p}_7 &=& 
(\DD_2\cdot\erre_1)\mathfrak{p}_{5}+
A\mathfrak{q}_6,
\end{eqnarray*}
where
\[
  \mathfrak{a}_1 = (\DD_1\cdot\erho_2)\,\mathfrak{f}_7 -
  \mathfrak{f}_4,\hskip 1cm
\mathfrak{a}_2 = -(\DD_2\cdot\erho_1)\,\mathfrak{f}_7 -
\mathfrak{f}_4.
\]
We observe that, using relations $\mathfrak{q}_2=\mathfrak{q}_3=0$ to
eliminate $\rhodot_1, \rhodot_2$ from
$\mathfrak{p}_6,\mathfrak{p}_7$, we obtain
\[
\tilde{\mathfrak{p}}_6 =
(-\DD_1\cdot\erho_2)p_1,\hskip 1cm
\tilde{\mathfrak{p}}_7 = 
(\DD_2\cdot\erho_1)p_2,
\]
where $p_1$, $p_2$ are the bivariate polynomials defined in (\ref{p1p2}).
Since we can write
\[
\tilde{\mathfrak{p}}_6 = \mathfrak{p}_6 + \mathfrak{b}_2\mathfrak{q}_2
+ \mathfrak{b}_3\mathfrak{q}_3,
\hskip 1cm \tilde{\mathfrak{p}}_7 = \mathfrak{p}_7 +
\mathfrak{c}_2\mathfrak{q}_2 + \mathfrak{c}_3\mathfrak{q}_3
\]
for some polynomials $\mathfrak{b}_j,\mathfrak{c}_j$, $j=2,3$ in the
variables $\rho_1,\rho_2,\rhodot_1,\rhodot_2$, we have
$\tilde{\mathfrak{p}}_6,\tilde{\mathfrak{p}}_7\in I$.

Let us consider the elimination ideal
\[
J = \langle\mathfrak{q}_1,\tilde{\mathfrak{p}}_6,\tilde{\mathfrak{p}}_7\rangle
= \langle q,p_1,p_2\rangle, 
\]
in $\R[\rho_1,\rho_2]$.
The ideal
\[
\tilde{J} = \langle q, \tilde{p}_1, \tilde{p}_2 \rangle,
\]
with the polynomials $\tilde{p}_j$ defined in (\ref{pjtilde}), coincides with
$J$, in fact
\[
\tilde{p}_j = p_j + d_jq, \hskip 1cm j=1,2
\]
for some polynomials $d_j = d_j(\rho_1,\rho_2)$. In particular, we have
\[
V(J)= V(\tilde{J}),
\]
where the variety $V(K)$ of a polynomial ideal $K\in\R[\rho_1,\rho_2]$ is
the set
\[
V(K) = \{(\rho_1,\rho_2)\in\C^2:p(\rho_1,\rho_2)=0,\ \forall p\in K\}.
\]
The ideal 
\[
\tilde{J}_1 = \langle\tilde{p}_1,\tilde{p}_2\rangle, 
\]
fulfills 
\begin{equation}
\tilde{J}_1 \subseteq \tilde{J},
\label{J1inJ}
\end{equation}
so that 
\begin{equation}
V(\tilde{J}_1)\supseteq V(\tilde{J}).
\label{VJinVJ1}
\end{equation}
Indeed, we shall show that
\[
V(\tilde{J}_1) = V(\tilde{J}).
\]
Let us introduce the polynomial 
\[
\newu := \mathrm{Res}(\tilde{p}_1,\tilde{p}_2,\rho_1) =
\tilde{a}_{1,1}\tilde{a}_{2,0} - \tilde{a}_{1,0}\tilde{a}_{2,1}.
\]
We need the following results.
\begin{lemma}
  For a generic choice of the data $\Att_j,\bq_j,\bqdot_j, j=1,2$, the
  polynomials $\oldu, \newu$ have 9 distinct solutions in $\C$. 
\label{ninesol}
\end{lemma}

{\em Proof.}
We show this property for $\oldu$; the proof for $\newu$ is analogous.
Let
\[
\oldu(\rho_2) = \sum_{j=0}^9 c_j\rho_2^j,
\]  
for some coefficients $c_j\in\R$ depending on the data.  First we show that, for a generic choice of
the data, the rank of the Jacobian matrix
\[
\frac{\partial (c_0,\ldots, c_9)}{\partial (\Att_1,\Att_2,\bq_1,\bqdot_1,\bq_2,\bqdot_2)}
\]
is maximal, that is equal to 10.
To check this property it suffices to show that this rank is maximal for a
particular choice of the data.  In fact, if the rank were $<10$ in an open
set,
then by the analytic dependence of the coefficients $c_j$ on the data the rank
would not be maximal at any point.
We made this check using the symbolic computation software
{\em Maple 18} 
with the following data:
\begin{eqnarray*}
&&\Att_1 = \Bigl(2\arctan\frac{1}{2},0,1,1\Bigr),\qquad
\Att_2 = \Bigl(2\arctan\frac{1}{2},2\arctan\frac{1}{2},1,1\Bigr), \\
&&\bq_1 = (1,0,0),\qquad\quad \bqdot_1 = (0,1,0),\qquad\quad
\bq_2 = (0,1,0),\qquad\quad \bqdot_2 = (-1,0,0).
\end{eqnarray*}
%
Moreover, by a well known property of polynomials, we know that
$\oldu$ is {\em square-free} (i.e. without multiple roots) for a
generic choice of the coefficients $c_j$. This fact, together with the maximal
rank property showed above, concludes the proof of the lemma.

\rightline{$\square$}

\begin{lemma}
For a generic choice of the data $\Att_j,\bq_j,\bqdot_j, j=1,2$, we have
\begin{equation}
  \mathrm{gcd}(\tilde{a}_{1,1},\tilde{a}_{2,1}) = 1,
\label{gcduno}
\end{equation}
where $\tilde{a}_{1,1}$, $\tilde{a}_{2,1}$ are the univariate polynomials
defined in (\ref{atilde1}), (\ref{atilde2}).
\label{l:relprime}
\end{lemma}

{\em Proof.}
  We give a proof similar to the one of Lemma~\ref{ninesol}.  
Let
\[
\tilde{a}_{11}(\rho_2) = \sum_{j=0}^3 c_{1,j}\rho_2^j, \hskip 1cm
\tilde{a}_{21}(\rho_2) = \sum_{j=0}^4 c_{2,j}\rho_2^j,
\]
for some coefficients  $c_{i,j}$ depending on the data.
We can show that
  the Jacobian matrix
\[
\frac{\partial (c_{1,0},\ldots,c_{1,3},c_{2,0},\ldots,c_{2,4})}{\partial
  (\Att_1,\Att_2,\bq_1,\bqdot_1,\bq_2,\bqdot_2)}
\]
has generically maximal rank, i.e. 9, by checking that the rank is maximal
for the data of Lemma~\ref{ninesol}.  To conclude we use the fact that for a
generic choice of the coefficients $c_{i,j}$ relation (\ref{gcduno}) holds
true.

\rightline{$\square$}
%
By Lemma~\ref{l:relprime} we can find two univariate polynomials $\beta,
\gamma$ in the variable $\rho_2$ such that
\begin{equation}
\beta\tilde{a}_{1,1} + \gamma\tilde{a}_{2,1} = 1.
\label{bezout}
\end{equation}
Let us introduce
\begin{equation}
  \mathfrak{w} := \beta\tilde{p}_1 + \gamma\tilde{p}_2 = 
  \rho_1 + \mathfrak{z}(\rho_2),
\label{vudoppio}
\end{equation}
where
\[
\mathfrak{z} = \beta\tilde{a}_{1,0} + \gamma\tilde{a}_{2,0}.
\]
\begin{lemma}
The polynomial ideal
\[
\tilde{J}_2 = \langle \mathfrak{w}, \newu\rangle
\]
is equal to $\tilde{J}_1$.
\label{J1J2}
\end{lemma}

{\em Proof.}
From the definition of $\mathfrak{w}$ 
and from relation
\begin{equation}
\newu = \tilde{a}_{1,1}\tilde{p}_2 - \tilde{a}_{2,1}\tilde{p}_1  
\label{hatu}
\end{equation}
we have $\tilde{J}_2\subseteq \tilde{J}_1$.  On the other hand, we can easily
invert relations (\ref{vudoppio}), (\ref{hatu}) and, using (\ref{bezout}), we
obtain
\[
\tilde{p}_1 = \tilde{a}_{1,1}\mathfrak{w} + \gamma \newu,
\hskip 1cm
\tilde{p}_2 = \tilde{a}_{2,1}\mathfrak{w} - \beta \newu,
\] 
so that the other inclusion $\tilde{J}_1\subseteq \tilde{J}_2$ holds true.

\rightline{$\square$}
Lemmata~\ref{ninesol}, \ref{J1J2} imply  that $V(\tilde{J}_1)$ has
9 distinct points.
In fact, from $\mathfrak{w}=0$, for each root $\rho_2$ of $\newu$
we find a unique $\rho_1$ such that $(\rho_1,\rho_2)\in V(\tilde{J}_1)$.
On the other hand, since $\tilde{J}=J$ we have $V(\tilde{J})=V(J)$ and
generically $V(J)$ has 9 distinct points too. We can prove it by using Theorem
1 in \cite{gbm15} and Lemma~\ref{ninesol} for the polynomial $\oldu$.
%
%
Then from (\ref{VJinVJ1}) we conclude that
\begin{equation}
V(\tilde{J}_1) = V(\tilde{J}).
\label{samevar}
\end{equation}
In particular, the polynomials $\newu$ and $\oldu$ coincide up to a constant
factor.

Now we prove that $\tilde{J}_1$ is indeed equal to $\tilde{J}$.
Let us take $h\in \tilde{J}$. Making the division by $\mathfrak{w}$ we
obtain
\begin{equation}
h(\rho_1,\rho_2) = h_1(\rho_1,\rho_2)\bigl(\rho_1 +
\mathfrak{z}(\rho_2)\bigr) + \mathfrak{r}(\rho_2)
\label{decomp}
\end{equation}
for some polynomials $h_1,\mathfrak{r}$. The remainder $\mathfrak{r}$ depends
only on $\rho_2$ because $\mathfrak{w}$ is linear in $\rho_1$.  From
(\ref{J1inJ}) and (\ref{decomp}) we have that
$\mathfrak{r}\in\tilde{J}$. Using relation (\ref{samevar}) and the fact that
$\oldu$ is generically squarefree we obtain
\[
\oldu \mid \mathfrak{r}, 
\]
that together with (\ref{decomp}) implies that $h\in\tilde{J}_1$.
We conclude that 
\[
\tilde{J}_1 = \tilde{J}.
\]

The polynomials $\mathfrak{g}_1\ldots\mathfrak{g}_6$, with
\[
\mathfrak{g}_1 = \mathfrak{q}_2,\qquad
\mathfrak{g}_2 = \mathfrak{q}_3,\qquad
\mathfrak{g}_3 = \mathfrak{p}_4,\qquad
\mathfrak{g}_4 = \mathfrak{p}_5,\qquad
\mathfrak{g}_5 = \mathfrak{w},\qquad
\mathfrak{g}_6 = \mathfrak{u},
\]
form a Groebner basis of the ideal $I$ for the lexicographic order
(\ref{lexord}).  To show this we can simply check that the leading monomials
of each pair ($\mathfrak{g}_i, \mathfrak{g}_j$), with $1\leq i<j\leq 6$, are
relatively prime (see \cite{cox}, Chapter 2).
This concludes the proof of the theorem.

\rightline{$\square$}

From the definition of Groebner basis we immediately obtain the following
\begin{corollary}
The polynomial $\oldu$ has the least degree among the univariate
polynomials in the variable $\rho_2$ belonging to the ideal $I$. 
\end{corollary}

As a consequence of the computations in the proof of Theorem~\ref{mainteo} we
also obtain
\begin{corollary}
  The polynomial system~(\ref{fullsys}) is generically not consistent. The
  same result holds true by removing from (\ref{fullsys}) only one of the two
  equations $u^2_j|\erre_j|^2 = \mu^2$, $j=1,2$.
\label{inconsist}
\end{corollary}
{\em Proof.}
We show that the system
\begin{equation}
\mathfrak{g}_j = 0, \ \ j=1\ldots 6, \qquad u^2_2|\erre_2|^2 - \mu^2 = 0
\label{intermedsys}
\end{equation}
is generically not consistent, where $\mathfrak{g}_j$ are the polynomials in
the statement of Theorem~\ref{mainteo}.  By using equations 
$\mathfrak{g}_1 = \mathfrak{g}_2 = \mathfrak{g}_3 = \mathfrak{g}_5 = 0$
we can obtain from
$u^2_2|\erre_2|^2 = \mu^2$ another univariate polynomial, say
$\hat{\mathfrak{u}}$ in the variable $\rho_2$.  Then $\mathfrak{u}$ and
$\hat{\mathfrak{u}}$ have a common root in $\C$ (i.e. are compatible) if and
only if
\begin{equation}
\mathrm{Res}(\mathfrak{u}, \hat{\mathfrak{u}},\rho_2)=0.
\label{resuhatu}
\end{equation} 
Assume there is an open set in the space of the data $\Att_j, \bq_j, \bqdot_j,
j=1,2$ such that equation (\ref{resuhatu}) holds.  Since the left-hand side of
(\ref{resuhatu}) is an analytic function of the data, then this equation holds
on the whole data set.  Therefore, to conclude it is enough to check that
equations (\ref{intermedsys}) are not compatible for a particular choice of
the data, e.g. as in Lemma~\ref{ninesol}.

In a similar way we can prove that the system
\[
\mathfrak{g}_j = 0, \ \ j=1\ldots 6, \qquad u^2_1|\erre_1|^2 - \mu^2 = 0
\]
is generically not consistent.

\rightline{$\square$}

\subsection{Compatibility conditions and covariance of the solutions}
\label{s:comptwo}

In this section we discuss how to discard some of the solutions computed with
the method described in Section~\ref{s:linktwo} on the base of the full
two-body dynamics.
Given a pair of attributables $\Avec = (\Att_1,\Att_2)$ at epochs $\bar{t}_1,
\bar{t}_2$ with covariance matrices $\Gamma_{\Att_1}, \Gamma_{\Att_2}$, we
call $\Rvec = (\rho_1, \rhodot_1, \rho_2, \rhodot_2)$ one of the solutions of
the equation
\begin{equation}
\bPhi(\Rvec;\Avec) = {\bf 0},
\label{Phi2_eq_0}
\end{equation}
with
\[
\bPhi(\Rvec; \Avec) = \left(
\begin{array}{c}
\angmom_1 - \angmom_2\cr
\bXi\cdot\erho_1
\end{array}
\right),
\]
where
\[
\bXi =
\frac{1}{2}(|\erredot_2|^2- |\erredot_1|^2)\erre_1\times\erre_2
- (\erredot_1\cdot\erre_1)\erredot_1\times(\erre_1-\erre_2) +
(\erredot_2\cdot\erre_2)\erredot_2\times(\erre_1-\erre_2),
\]
which corresponds to the vector $\bxi$ defined in (\ref{bxi}) if we eliminate
$\rhodot_1, \rhodot_2$ by (\ref{rhojdot}).
We can repeat what follows for each solution of $\bPhi(\Rvec;\Avec) = {\bf
  0}$.  The notation is similar to \cite{gdm10}.

Let us introduce the difference vector
\[
\bDelta_{a,\ell} = (\Delta_a,\Delta\ell),
\]
where 
\[
\Delta_a = a_1-a_2, \hskip 1cm \Delta_\ell = \bigl[\ell_1-\bigl(\ell_2 +
n(a_2)(\tilde{t}_1-\tilde{t}_2)\bigr)+\pi (\mathrm{mod}\ 2\pi)\bigr] -\pi,
\]
where $n(a) = \sqrt{\mu} a^{-3/2}$ is the mean motion and $\tilde{t}_i =
\bar{t}_i - \rho_i/c$, $i=1,2$.  Note that here we consider the difference of
the two mean anomalies at the same epoch $\tilde{t}_1$ in a way that it is a
smooth function at each integer multiple of $2\pi$.
We introduce the map
\[
(\Att_1, \Att_2) = \Avec \mapsto \bPsi(\Avec) = \left(\Att_1,
  \Rcal_1,\bDelta_{a,\ell}\right),
\]
giving the orbit $(\Att_1, \Rcal_1)$ in attributables coordinates at epoch
$\tilde{t}_1$ together with the vector $\bDelta_{a,\ell}$ which is not
constrained by equation (\ref{Phi2_eq_0}).

By the covariance propagation rule we have
\begin{equation}
\Gamma_{\bPsi(\Avec)} = \frac{\partial \bPsi}{\partial
\Avec }\; \Gamma_\Avec \; \left[\frac{\partial \bPsi}{\partial
\Avec }\right]^T\ ,
\label{cov_prop_rule}
\end{equation}
where
\[
 \frac{\partial
\bPsi}{\partial \Avec} = \left[
\begin{array}{ccc}
I &0\cr
 \displaystyle\frac{\partial \Rcal_1}{\partial \Att_1} 
&\displaystyle\frac{\partial \Rcal_1}{\partial \Att_2} \cr 
\stackrel{}{\displaystyle\frac{\partial\bDelta_{a,\ell}}{\partial \Att_1}} 
&\displaystyle\frac{\partial\bDelta_{a,\ell}}{\partial \Att_2} \cr 
\end{array}
\right]
\hskip 0.5cm\mbox{ and }
\hskip 0.5cm
\Gamma_\Avec =  \left[
\begin{array}{cc}
\Gamma_{\Att_1} &0 \cr
0 &\Gamma_{\Att_2} \cr
\end{array}
\right] .
\]
We can check if there is any solution of (\ref{Phi2_eq_0})
fulfilling the compatibility conditions
\[
\bDelta_{a,\ell} = \bzero
\]
within a threshold defined by the covariance matrix of the attributables
$\Gamma_\Avec$.
From (\ref{cov_prop_rule}) we can compute the marginal covariance of the
vector $\bDelta_{a,\ell}$:
\[
\Gamma_{\bDelta_{a,\ell}} = 
\frac{\partial \bDelta_{a,\ell}}{\partial\Avec}\Gamma_\Avec 
\left[\frac{\partial \bDelta_{a,\ell}}{\partial\Avec}\right]^T.
\]
The inverse matrix $C^{\bDelta_{a,\ell}} = \Gamma^{-1}_{\bDelta_{a,\ell}}$
defines a norm $\Vert\cdot\Vert_{\star}$ in the $(\Delta a, \Delta\ell)$
plane, allowing us to test an identification between the attributables
$\Att_1, \Att_2$: we check whether
\[
\Vert \bDelta_{a,\ell}\Vert_{\star}^2 = \bDelta_{a,\ell} C^{\bDelta_{a,\ell}}
\bDelta_{a,\ell}^T\leq \chi_{max}^2,
\]
where $\chi_{max}$ is a control parameter.

If a preliminary orbit $(\Att_1, \Rcal_1)$ is accepted, from
(\ref{cov_prop_rule}) we can also compute its marginal covariance as the
$6\times 6$ matrix
\[
\Gamma_{(\Att_1, \Rcal_1)}
=\left[ 
\begin{array}{ccc}
\Gamma_{\Att_1}  
&\Gamma_{\Att_1,\Rcal_1}\cr
\Gamma_{\Rcal_1,\Att_1}
&\Gamma_{\Rcal_1}\cr
\end{array}
\right],
\]
%
where
\[
\Gamma_{\Att_1,\Rcal_1} = \Gamma_{\Att_1}\left[\frac{\partial
\Rcal_1}{\partial\Att_1}\right]^T,
\qquad
\Gamma_{\Rcal_1} = \frac{\partial \Rcal_1}{\partial\Avec}\Gamma_\Avec 
\left[\frac{\partial \Rcal_1}{\partial\Avec}\right]^T,
\qquad
\Gamma_{\Rcal_1,\Att_1} = \Gamma_{\Att_1,\Rcal_1}^T.
\]

\section{Linking three VSAs}
\label{s:linkthree}

Here we introduce a method to compute preliminary orbits from three VSAs
using the Keplerian integrals (\ref{kepint}).
In this case the conservation of the angular momentum at the three
epochs is enough to obtain a finite number of solutions of the identification
problem.
In the following the indexes $1,2,3$ will refer to the mean epochs $\bar{t}_j$
of three VSAs with attributables ${\cal A}_j$.
We consider the equations:
\begin{equation}
\angmom_1 = \angmom_2,
\qquad
\angmom_2 = \angmom_3,
\qquad
\angmom_3 = \angmom_1,
\label{angmomeqs}
\end{equation}
that can be written as
\[
\DD_1\rhodot_1 - \DD_2\rhodot_2 = \JJ_{12}(\rho_1,\rho_2),\hskip 0.5cm
%
\DD_2\rhodot_2 - \DD_3\rhodot_3 = \JJ_{23}(\rho_2,\rho_3),\hskip 0.5cm
%
\DD_3\rhodot_3 - \DD_1\rhodot_1 = \JJ_{31}(\rho_3,\rho_1),
%
\]
where
\begin{eqnarray*}
\JJ_{12}(\rho_1,\rho_2) &=& \EE_2\rho_2^2 - \EE_1\rho_1^2 + \FF_2\rho_2 -
\FF_1\rho_1 + \GG_2 - \GG_1,\\
\JJ_{23}(\rho_2,\rho_3) &=& \EE_3\rho_3^2 - \EE_2\rho_2^2 + \FF_3\rho_3 -
\FF_2\rho_2 + \GG_3 - \GG_2,\\
\JJ_{31}(\rho_3,\rho_1) &=& \EE_1\rho_1^2 - \EE_3\rho_3^2 + \FF_1\rho_1 -
\FF_3\rho_3 + \GG_1 - \GG_3.
\end{eqnarray*}
Equations (\ref{angmomeqs}) are redundant, that is, if two of them
hold true then the third equation is also fulfilled.
We consider the following projections of equations (\ref{angmomeqs}):
\begin{eqnarray}
(\angmom_1-\angmom_2)\cdot\DD_1\times\DD_2 = 0,\label{AM12uno}\\
(\angmom_1-\angmom_2)\cdot\DD_1\times(\DD_1\times\DD_2) = 0,\label{AM12due}\\
(\angmom_2-\angmom_3)\cdot\DD_2\times\DD_3 = 0,\label{AM23uno}\\
(\angmom_2-\angmom_3)\cdot\DD_2\times(\DD_2\times\DD_3) = 0,\label{AM23due}\\
(\angmom_3-\angmom_1)\cdot\DD_3\times\DD_1 = 0,\label{AM31uno}\\
(\angmom_3-\angmom_1)\cdot\DD_3\times(\DD_3\times\DD_1) = 0.\label{AM31due}
\end{eqnarray}

\begin{proposition}
Assume 
\begin{equation}
\DD_1\times\DD_2\cdot\DD_3 \neq 0.
\label{D1D2D3}
\end{equation}
Then the system of equations (\ref{AM12uno})--(\ref{AM31due}) is
equivalent to (\ref{angmomeqs}).
\end{proposition}
{\em Proof.}
Assuming that (\ref{AM31uno}), (\ref{AM31due}) are fulfilled, to prove
that $\angmom_3=\angmom_1$ we only need to show that the projection of
this equation onto a vector $\bv$, such that
$\DD_3\times\DD_1, \DD_3\times(\DD_3\times\DD_1),\bv$ are linearly
independent, holds true.  We denote by 
\[
\Pi_{12} =
\langle\DD_1\times\DD_2, \DD_1\times(\DD_1\times\DD_2) \rangle,
\hskip 1cm
\Pi_{23} =
\langle\DD_2\times\DD_3, \DD_2\times(\DD_2\times\DD_3) \rangle
\]
the planes passing through the origin generated by the vectors within the
brackets.  If relation (\ref{D1D2D3}) holds, then we have
\[
\Pi_{12}\cap\Pi_{23} = \langle\DD_1\times\DD_2\rangle,
\]
i.e. the intersection of the two planes is the straight line
generated by the vector $\bv = \DD_1\times\DD_2$.
Moreover, we have
\[
(\DD_1\times\DD_2) \cdot 
(\DD_3\times\DD_1)\times\bigl(\DD_3\times(\DD_3\times\DD_1)\bigr) = 
|\DD_3\times\DD_1|^2 \DD_1\times\DD_2\cdot\DD_3,
\]
that does not vanish by (\ref{D1D2D3}).
Therefore, from (\ref{AM12uno})--(\ref{AM23due}) we obtain
$(\angmom_1-\angmom_2)\cdot\bv = (\angmom_2-\angmom_3)\cdot\bv = 0$, that
yield $(\angmom_3-\angmom_1)\cdot\bv = 0$. 
In a similar way we can prove that $\angmom_1=\angmom_2$,
$\angmom_2=\angmom_3$, provided (\ref{AM12uno})--(\ref{AM31due})
hold.

\rightline{$\square$}

Equations (\ref{AM12uno}), (\ref{AM23uno}), (\ref{AM31uno})
depend only on the radial distances. In fact, they correspond to the system
\begin{equation}
\JJ_{12}\cdot \DD_1\times\DD_2 = 0,
\qquad
\JJ_{23}\cdot \DD_2\times\DD_3 = 0,
\qquad
\JJ_{31}\cdot \DD_3\times\DD_1 = 0,
\label{JJ123}
\end{equation}
which can be written as
\begin{eqnarray}
q_3 &=& a_3\rho_2^2 + b_3\rho_1^2 + c_3\rho_2 + d_3\rho_1 + e_3 = 0,
\label{qtre}\\
q_1 &=& a_1\rho_3^2 + b_1\rho_2^2 + c_1\rho_3 + d_1\rho_2 + e_1 = 0,
\label{quno}\\
q_2 &=& a_2\rho_1^2 + b_2\rho_3^2 + c_2\rho_1 + d_2\rho_3 + e_2 = 0,\label{qdue}
\end{eqnarray}
where
\begin{eqnarray*}
&&a_3 =  \EE_2\cdot \DD_1\times\DD_2,\qquad
b_3 = -\EE_1\cdot \DD_1\times\DD_2,\\
&&c_3 =  \FF_2\cdot \DD_1\times\DD_2,\qquad
d_3 = -\FF_1\cdot \DD_1\times\DD_2,\\
&&\hskip 1cm e_3 =  (\GG_2-\GG_1)\cdot \DD_1\times\DD_2,
\end{eqnarray*}
and the other coefficients $a_j, b_j, c_j, d_j, e_j$, for $j=1,2$, have
similar expressions, obtained by cycling the indexes.
To eliminate $\rho_1, \rho_3$ from (\ref{JJ123}) we first compute the
resultant
\[
r = \mathrm{Res}(q_3,q_2,\rho_1),
\]
which depends only on $\rho_2,\rho_3$. Then we compute the resultant
\[
\mathfrak{q} = \mathrm{Res}(r,q_1,\rho_3),
\]
which is a univariate polynomial of degree 8 in the variable $\rho_2$.
Therefore, provided (\ref{D1D2D3}) holds, to get the solutions of
(\ref{angmomeqs}) we search for the roots $\bar{\rho}_2$ of
$\mathfrak{q}(\rho_2)$, then we compute the corresponding values $\bar{\rho}_3$
from system $r(\rho_3,\bar{\rho}_2) = q_1(\rho_3,\bar{\rho}_2) = 0$, and
finally the corresponding values $\bar{\rho}_1$ from system
$q_3(\rho_1,\bar{\rho}_2) = q_2(\bar{\rho}_3,\rho_1) = 0$.
Since the unknowns $\rho_j$ represent distances we can discard triples
$(\bar{\rho}_1,\bar{\rho}_2,\bar{\rho}_3)$ where some $\rho_j$ is
non-positive.
From equations (\ref{AM12due}), (\ref{AM23due}), (\ref{AM31due})
we can write the radial velocities $\rhodot_j$ as functions of pairs of
radial distances:
\begin{eqnarray*}
\rhodot_2 &=&
\frac{\JJ_{12}(\rho_1,\rho_2)\cdot\DD_1\times(\DD_1\times\DD_2)}{|\DD_1\times\DD_2|^2},\\
\rhodot_3 &=&
\frac{\JJ_{23}(\rho_2,\rho_3)\cdot\DD_2\times(\DD_2\times\DD_3)}{|\DD_2\times\DD_3|^2},\\
\rhodot_1 &=&
\frac{\JJ_{31}(\rho_3,\rho_1)\cdot\DD_3\times(\DD_3\times\DD_1)}{|\DD_3\times\DD_1|^2}.
\end{eqnarray*}

\begin{remark}
  As a simple criterion to discard triples $(\Att_1, \Att_2, \Att_3)$ before
  making the computation described in this section we can use the intersection
  criterion introduced in \cite{gbm15} to discard pairs of attributables. 
  More precisely, we can apply this criterion three times, i.e.  we check for
  each $j=1,2,3$ whether the conic $Q_j$, defined by $q_j=0$ (see equations
  (\ref{qtre}), (\ref{quno}), (\ref{qdue})), intersects the square ${\cal R} =
  [\rho_{min},\rho_{max}]\times[\rho_{min},\rho_{max}]$ for some fixed
  $\rho_{max}>\rho_{min}>0$.
  If this criterion fails in one of these cases we discard the selected
  triple.  For more details see the appendix in \cite{gbm15}.
\label{r:quickdiscard}
\end{remark}

\subsection{Solutions with zero angular momentum}

A particular solution of system (\ref{angmomeqs}) can be obtained by searching
for values of $\rho_j,\rhodot_j$ such that 
\[
\angmom_j(\rho_j,\rhodot_j) = \bzero, \hskip 1cm j=1,2,3.
\]
Relation $\erre\times\erredot = \bzero$ implies that there exists
$\lambda\in\R$ such that
\begin{equation}
\rhodot\erho + \rho\etabf + \bqdot = \lambda(\rho\erho + \bq),
\label{parall}
\end{equation}
with $\etabf = \alphadot\cos\delta\ealpha + \deltadot\edelta$.
Setting $\sigma = \rhodot-\lambda\rho$ we can write (\ref{parall}) as
\begin{equation}
\sigma\erho + \rho\etabf - \lambda\bq = -\bqdot.
\label{parall2}
\end{equation}
We introduce the vector
\[
\bu = \bq - (\bq\cdot\erho)\erho - \frac{1}{\eta^2}(\bq\cdot\etabf)\etabf,
\]
which is orthogonal to both $\erho, \etabf$, where $\eta = |\etabf|$ is called
the {\em proper motion}.
Thus, we can write (\ref{parall2}) as
\[
[\sigma - \lambda(\bq\cdot\erho)]\erho +
\Bigl[\rho-\frac{\lambda}{\eta^2}(\bq\cdot\etabf)\Bigr]\etabf - \lambda\bu =
-\bqdot.
\]
Since $\{\erho,\etabf,\bu\}$ is generically an orthogonal basis of $\R^3$, we
find
\[
\lambda = \frac{1}{|\bu|^2}(\bqdot\cdot\bu),\qquad
\rho = \frac{1}{\eta^2}(\lambda\bq - \bqdot)\cdot\etabf,\qquad
\rhodot = \lambda\rho + (\lambda\bq - \bqdot)\cdot\erho.
\]

\noindent In particular we obtain the value 
\[
  \rho = \frac{1}{\eta^2}
  \Bigl(\frac{1}{|\bu|^2}(\bqdot\cdot\bu)(\bq\cdot\etabf) - 
    \bqdot\cdot\etabf\Bigr)
\]
for the radial distance, corresponding to a solution with zero angular
momentum.

\subsection{Compatibility conditions and covariance of the solutions}
\label{s:compthree}

We discuss how to discard solutions of (\ref{angmomeqs}) in a way similar to
Section~\ref{s:comptwo}.
Given a triple of attributables $\Avec = (\Att_1,\Att_2,\Att_3)$ with
covariance matrices $\Gamma_{\Att_1}, \Gamma_{\Att_2}, \Gamma_{\Att_3}$, we
call $\Rvec = (\rho_1, \rhodot_1, \rho_2, \rhodot_2, \rho_3, \rhodot_3)$ one
of the solutions of the equation 
\begin{equation}
\bPhi(\Rvec;\Avec) = {\bf 0},
\label{Phi_eq_0}
\end{equation}
with
\[
\bPhi(\Rvec; \Avec) = \left(
\begin{array}{c}
(\angmom_1 - \angmom_2)\cdot\DD_1\times(\DD_1\times\DD_2)\cr
(\angmom_1 - \angmom_2)\cdot\DD_1\times\DD_2\cr
(\angmom_2 - \angmom_3)\cdot\DD_2\times(\DD_2\times\DD_3)\cr
(\angmom_2 - \angmom_3)\cdot\DD_2\times\DD_3\cr
(\angmom_3 - \angmom_1)\cdot\DD_3\times(\DD_3\times\DD_1)\cr
(\angmom_3 - \angmom_1)\cdot\DD_3\times\DD_1\cr
\end{array}
\right) .
\]
We can repeat what follows for each solution of $\bPhi(\Rvec;\Avec) = {\bf 0}$.

%
Let us introduce the difference vectors
\begin{eqnarray*}
\bDelta_{12} &=& \bigl(a_1-a_2, 
[\omega_1-\omega_2 +\pi (\mathrm{mod}\ 2\pi)] -\pi, 
\bigl[\ell_1-\bigl(\ell_2 + n(a_2)(\tilde{t}_1-\tilde{t}_2)\bigr) +\pi
(\mathrm{mod}\ 2\pi)\bigr] -\pi \bigr),\\
\bDelta_{32} &=&\bigl(a_3-a_2, 
[\omega_3-\omega_2+\pi (\mathrm{mod}\ 2\pi)] -\pi, 
\bigl[\ell_3-\bigl(\ell_2 + n(a_2)(\tilde{t}_3-\tilde{t}_2)\bigr) +\pi 
(\mathrm{mod}\ 2\pi)\bigr] -\pi \bigr),
\end{eqnarray*}
where the third component is the difference of the two mean anomalies
referring to epoch $\tilde{t}_i = \bar{t}_i - \rho_i/c$,
and $n(a) = \sqrt{\mu} a^{-3/2}$ is the mean motion.
Here the difference of two angles is computed in a way that it is a smooth
function at each integer multiple of $2\pi$.
We introduce the map
\[
(\Att_1, \Att_2 ,\Att_3) = \Avec \mapsto \bPsi(\Avec) = \left(\Att_2,
  \Rcal_2,\bDelta_{12}, \bDelta_{32}\right),
\]
giving the orbit $(\Att_2,\Rcal_2(\Avec))$ in attributable coordinates at epoch
$\tilde{t}_2$ together with the vectors $\bDelta_{12}(\Avec)$,
$\bDelta_{32}(\Avec)$, which are not constrained by the angular momentum
integrals.  We want to check if there is any solution of (\ref{Phi_eq_0})
fulfilling the compatibility conditions
\[
\bDelta_{12} = \bDelta_{32} = \bzero
\]
within a threshold defined by the covariance matrix of the attributables
\[
\Gamma_\Avec =  \left[
\begin{array}{ccc}
\Gamma_{\Att_1} &0 &0\cr
0 &\Gamma_{\Att_2} &0\cr
0 &0 &\Gamma_{\Att_3}\cr
\end{array}
\right].
\]
By the covariance propagation rule we have
\[
\Gamma_{\bPsi(\Avec)} = \frac{\partial \bPsi}{\partial
\Avec }\; \Gamma_\Avec \; \left[\frac{\partial \bPsi}{\partial
\Avec }\right]^T\ ,
\]
where
\[
 \frac{\partial
\bPsi}{\partial \Avec} = \left[
\begin{array}{ccc}
0 &I &0\cr
 \displaystyle\frac{\partial \Rcal_2}{\partial \Att_1} 
&\displaystyle\frac{\partial \Rcal_2}{\partial \Att_2} 
&\displaystyle\frac{\partial \Rcal_2}{\partial \Att_3} \cr
\stackrel{}{\displaystyle\frac{\partial\bDelta_{12}}{\partial \Att_1}} 
&\displaystyle\frac{\partial\bDelta_{12}}{\partial \Att_2} 
&\displaystyle\frac{\partial\bDelta_{12}}{\partial \Att_3} \cr
\stackrel{}{\displaystyle\frac{\partial\bDelta_{32}}{\partial \Att_1}} 
&\displaystyle\frac{\partial\bDelta_{32}}{\partial \Att_2} 
&\displaystyle\frac{\partial\bDelta_{32}}{\partial \Att_3} \cr
\end{array}
\right].
\]
The matrices $\frac{\partial \Rcal_2}{\partial\Att_j}, j=1,2,3$, can be
computed from the relation
\[
\frac{\partial \Rvec}{\partial \Avec }(\Avec) = - \left[ \frac{\partial
\bPhi}{\partial \Rvec}(\Rvec(\Avec),\Avec) \right]^{-1}
\frac{\partial \bPhi}{\partial\Avec}(\Rvec(\Avec),\Avec) .
\]
The marginal covariance matrix for the vector $(\bDelta_{12}, \bDelta_{32})$
is given by the block
\[
\Gamma_{\bDelta}
=\left[ 
\begin{array}{cc}
\Gamma_{\bDelta_{12}}
&\Gamma_{\bDelta_{12},\bDelta_{32}}\cr
\Gamma_{\bDelta_{32},\bDelta_{12}}
&\Gamma_{\bDelta_{32}}\cr
\end{array}
\right]
\]
of $\Gamma_{\bPsi(\Avec)}$, 
where
\[
\begin{array}{cl}
\displaystyle\Gamma_{\bDelta_{12}} = \frac{\partial \bDelta_{12}}{\partial\Avec}\Gamma_\Avec 
\left[\frac{\partial \bDelta_{12}}{\partial\Avec}\right]^T,
&\hskip 0.5cm \displaystyle\Gamma_{\bDelta_{12},\bDelta_{32}} = \frac{\partial \bDelta_{12}}{\partial\Avec}\Gamma_\Avec 
\left[\frac{\partial \bDelta_{32}}{\partial\Avec}\right]^T,\cr
\stackrel{}{\displaystyle\Gamma_{\bDelta_{32}} = 
\frac{\partial \bDelta_{32}}{\partial\Avec}\Gamma_\Avec 
\left[\frac{\partial \bDelta_{32}}{\partial\Avec}\right]^T,}
&\hskip 0.5cm\displaystyle\Gamma_{\bDelta_{32},\bDelta_{12}} =
\Gamma_{\bDelta_{12},\bDelta_{32}}^T. \cr
\end{array}
\]
The inverse matrix $C^{\bDelta} = \Gamma^{-1}_{\bDelta}$ defines a norm
$\Vert\cdot\Vert_{\star}$ in the six dimensional space with coordinates
$\bDelta = (\bDelta_{12}, \bDelta_{32})$, allowing us to test an
identification between the attributables $\Att_1, \Att_2, \Att_3$: we check
whether 
\begin{equation}
\Vert \bDelta\Vert_{\star}^2 = \bDelta C^{\bDelta}
\bDelta^T\leq \chi_{max}^2,
\label{control_3}
\end{equation}
where $\chi_{max}$ is a control parameter.

For each orbit, solution of (\ref{Phi_eq_0}), fulfilling condition
(\ref{control_3}) we can also define a covariance matrix $\Gamma_2$ for the
attributable coordinates $({\cal A}_2,{\cal R}_2)$:
\[
\Gamma_2 = \left[ 
\begin{array}{cc}
\Gamma_{{\cal A}_2}
&\Gamma_{{\cal A}_2,{\cal R}_2}\cr
\Gamma_{{\cal R}_2, {\cal A}_2}
&\Gamma_{{\cal R}_2}\cr
\end{array}
\right],
\]
where $\Gamma_{{\cal A}_2}$ is given and
\[
\Gamma_{\Att_2,\Rcal_2} = \Gamma_{\Att_2}\left[\frac{\partial
\Rcal_2}{\partial\Att_2}\right]^T,
\qquad
\Gamma_{\Rcal_2} = \frac{\partial \Rcal_2}{\partial\Avec}\Gamma_\Avec 
\left[\frac{\partial \Rcal_2}{\partial\Avec}\right]^T,
\qquad
\Gamma_{\Rcal_2,\Att_2} = \Gamma_{\Att_2,\Rcal_2}^T.
\]

\section{Numerical tests}
\label{s:numtests}

In this section we compare the preliminary orbits obtained by Gauss' method and
the methods described in Sections~\ref{s:linktwo}, \ref{s:linkthree} for a
test case: the near-Earth asteroid (154229).
%
In Table~\ref{tabobs} we list three tracklets, each composed by four
observations (right ascension, declination), of this asteroid collected with
the Pan-STARRS telescope.
\begin{table}[h!]
\begin{center}
\begin{tabular}{c|c|c|c|c}
tr&obs&$\alpha$ (rad) &$\delta$ (rad) &epoch (MJD)\cr
\hline
1 &1 &3.834760347106644 &-7.983116606074819E-02 &57052.58743759259 \cr
  &2 &3.834778963951999 &-7.982534829657487E-02 &57052.59951759259 \cr
  &3 &3.834797653519405 &-7.981962749513778E-02 &57052.61160759259 \cr
  &4 &3.834816924863230 &-7.981410061917313E-02 &57052.62370759259 \cr
\hline
2 &1 &3.717640827594138 &4.346887946196211E-03  &57102.52326759259 \cr
  &2 &3.717559015285451 &4.378546279572663E-03  &57102.53596759259 \cr
  &3 &3.717476330312138 &4.410543982525893E-03  &57102.54885759259 \cr
  &4 &3.717393936227034 &4.442250797270456E-03  &57102.56162759259 \cr
\hline
3 &1 &3.369239074975656 &7.801563578772919E-02  &57163.27290759259 \cr
  &2 &3.369201695840843 &7.800763636199089E-02  &57163.28720759259 \cr
  &3 &3.369164534872187 &7.800021871266992E-02  &57163.30153759259 \cr
  &4 &3.369126864849164 &7.799251017514028E-02  &57163.31588759259 \cr
\end{tabular}
\end{center}
\caption{The three selected tracklets, each composed by four observations of 
  asteroid (154229).}
\label{tabobs}
\end{table}

\noindent In Table~\ref{tabatt} we show the approximated values of the
components of the three attributables computed from the tracklets in
Table~\ref{tabobs}.
\begin{table}[h!]
\begin{center}
\begin{tabular}{c|c|c|c|c|c}
att &$\alpha$ (rad) &$\delta$ (rad) &$\alphadot$ (rad/s) &$\deltadot$ (rad/s) &epoch (MJD)\cr
\hline
1 &3.83479 &-7.98225E-02 &\phantom{-}1.55849E-03 
  &\phantom{-}4.70783E-04 &57052.60557 \cr
2 &3.71752 &\phantom{-}4.39460E-03 &-6.43398E-03 
  &\phantom{-}2.48563E-03 &57102.54243 \cr
3 &3.36918 &\phantom{-}7.80039E-02 &-2.60900E-03 
  &-5.36020E-04 &57163.29439 \cr
\end{tabular}
\end{center}
\caption{Attributables computed from the three tracklets in Table~\ref{tabobs}.}
\label{tabatt}
\end{table}

\noindent To compare the preliminary orbits we use two least squares
solutions: one is computed with tracklets $1,2$ only, the other with all the
tracklets.  In Table~\ref{tabsolprel} we list these solutions together with
the preliminary orbits computed by the different methods.

\begin{table}[h!]
\begin{center}
\begin{tabular}{c|c|c|c|c|c|c|c|c}
        &epoch &$a \mathrm{\ (au)}$  &$e$  &$I$  &$\Omega$  &$\omega$  &$\ell$ &norm \cr
\hline 
$G_2$  &57077.57400 &1.85046 &0.71629 &10.00603 &66.70400 &343.12690 &59.78415 &4639.4 \cr
$L_2$  &57077.57400 &1.85384 &0.71913 &10.11799 &67.29283 &341.93359 &61.35804 &521.3 \cr
$LS_2$ &57077.57400 &1.84903 &0.71930 &10.09292 &67.65173 &341.39098 &61.73660 &// \cr
\hline
$G_3$  &57106.14746 &1.88095 &0.73082 &10.02343 &67.97447 &341.61797 &69.37321 &5882.0\cr
$L_3$  &57106.14746 &1.84725 &0.72153 &10.17272 &67.25235 &341.51657 &73.17327 &775.9\cr
$LS_3$ &57106.14746 &1.85112 &0.71865 &10.07393 &67.70983 &341.48650 &72.68650 &//\cr
\hline
$LS_2$ &57106.14746 &1.84899 &0.71930 &10.09304 &67.65083 &341.39054 &72.93972 &206660.0\cr
\end{tabular}
\end{center}
\caption{Preliminary orbits obtained with Gauss' method and with the linkage
  methods described in Sections~\ref{s:linktwo}, \ref{s:linkthree}. The angles
  are given in degrees. The values of the norms defined in (\ref{norms}), (\ref{norms2}) are listed in the last column.}
\label{tabsolprel}
\end{table}

\noindent The labels $G_2$, $G_3$ refer to the orbits obtained with Gauss'
method using different observations from Table~\ref{tabobs}: for $G_2$ we use
observations $1,4$ of tracklet 1 and observation 1 of tracklet 2; for $G_3$ we
use observation 1 of each tracklet.  The labels $L_2$, $L_3$ refer to the
methods described in Sections~\ref{s:linktwo}, \ref{s:linkthree}.  For $L_2$
we use attributables $1,2$ listed in Table~\ref{tabatt}; for $L_3$ we use all
the attributables in this table.  The labels $LS_2$, $LS_3$ refer to the least
squares orbits computed from $G_2$, $G_3$ respectively. For $LS_2$ we use the
observations of tracklets $1,2$ only, for $LS_3$ we use all the
observations in Table~\ref{tabobs}.
Let ${\cal E}_{G_2}$, ${\cal E}_{L_2}$, ${\cal E}_{G_3}$, ${\cal E}_{L_3}$ be
the preliminary orbits computed with the different methods.  Moreover, let
${\cal E}_{LS_2}$, ${\cal E}_{LS_3}$ be the least squares orbits corresponding
to the different sets of data employed, and let $\Gamma_{LS_2}$,
$\Gamma_{LS_3}$ be the related covariance matrices.  All the preliminary
orbits are propagated to the mean epoch of the arc of observations used to
compute the least squares solution ${\cal E}_{LS_2}$ or ${\cal E}_{LS_3}$,
according to the index $2$ or $3$.
Then we consider the normal matrices $C_{LS_2}=\Gamma_{LS_2}^{-1}$,
$C_{LS_3}=\Gamma_{LS_3}^{-1}$ corresponding to ${\cal E}_{LS_2}$, ${\cal
  E}_{LS_3}$.
The norms displayed in Table~\ref{tabsolprel} are defined as 
\begin{eqnarray}
&&|{\cal E}_{G_2}| = \Delta_{G_2}\cdot C_{LS_2}\Delta_{G_2},\quad
|{\cal E}_{L_2}| = \Delta_{L_2}\cdot C_{LS_2}\Delta_{L_2},\nonumber\\
\label{norms}\\
&&
|{\cal E}_{G_3}| = \Delta_{G_3}\cdot C_{LS_3}\Delta_{G_3},\quad
|{\cal E}_{L_3}| = \Delta_{L_3}\cdot C_{LS_3}\Delta_{L_3},\nonumber
\end{eqnarray}
where
\[
\Delta_{G_2} ={\cal E}_{G_2}-{\cal E}_{LS_2}, \qquad
\Delta_{L_2} ={\cal E}_{L_2}-{\cal E}_{LS_2}, \qquad
\Delta_{G_3} ={\cal E}_{G_3}-{\cal E}_{LS_3}, \qquad
\Delta_{L_3} ={\cal E}_{L_3}-{\cal E}_{LS_3}.
\]
The orbit ${\cal E}_{LS_2}$ in the last raw of Table~\ref{tabsolprel} is the
least squares solution obtained with tracklets $1,2$ and propagated at the
mean epoch of the three tracklets.  The corresponding norm is given by
\begin{equation}
|{\cal E}_{LS_2}| = \Delta_{LS_2}\cdot C_{LS_3}\Delta_{LS_2},
\label{norms2}
\end{equation}
where
\[
\Delta_{LS_2} ={\cal E}_{LS_2}-{\cal E}_{LS_3}.
\]
From the values of the norms in this test case we conclude that ${\cal
  E}_{L_3}$ is better than ${\cal E}_{G_3}$, because it is closer to
the least squares orbit ${\cal E}_{LS_3}$.  We also observe that ${\cal
  E}_{L_2}$ is better than ${\cal E}_{G_2}$.  However, the value of the
norm in the last raw in Table~\ref{tabsolprel} implies that ${\cal E}_{LS_2}$
is not close to ${\cal E}_{LS_3}$.
Both ${\cal E}_{L_3}$ and ${\cal E}_{G_3}$ are much better, as preliminary
orbits, than the orbit obtained by propagating ${\cal E}_{LS_2}$ to the
mean epoch of the three tracklets.
These results are consistent with the large scale test performed in
\cite[Fig. 2]{mgk07}, showing that least squares solutions with two VSAs are
very poor approximations of the true orbit, while least squares solutions with
three VSAs are accurate enough.
This implies the need to compute from scratch a preliminary orbit when we
join a third tracklet to a pair of linked VSAs.
The role of the linkage of two VSAs is to test the compatibility of pairs of
tracklets and discard a large number of them. The orbit computed with two VSAs
should not be used for the attribution (see \cite{ident4}) of a third
tracklet.

\section*{Acknowledgments}

We wish to thank M. Caboara, P. Gianni, E. Sbarra, B. Trager, who gave
us very useful suggestions on the algebraic aspects of this work.
This work is partially supported by the Marie Curie Initial Training
Network Stardust, FP7-PEOPLE-2012-ITN, Grant Agreement 317185.


\end{document}